\documentclass[a4paper]{jpconf}
\usepackage{graphicx}
\usepackage[english]{babel}
\usepackage{amssymb,amsmath}

\begin{document}
\title{Spiral structure of the Milky Way galaxy: observations and theoretical predictions}

\author{V. I. Korchagin$^{1*}$, S.S. Khrapov$^{2}$, A.V. Khoperskov$^{2}$}

\address{$^{1}$Southern Federal University, Rostov-on-Don, Russia \\

$^{2}$Volgograd State University, Volgograd, 400062, Russia
}

\ead{$^*$Corresponding author e-mail:  khoperskov@volsu.ru}

\begin{abstract}\footnotetext{
		VK acknowledges Russian Science Foundation (grant 18-12-00213) for financial support. AK and SK thank the Ministry of Science and Higher Education of the RF (project No. 2.852.2017/4.6) for the financial support of the software development.
		 {\bf \underline{Submitted}: Research in Astronomy and Astrophysics} 
	 }
Using observational data on the kinematical properties and density distributions of the subsystems
of the Milky Way galaxy, we construct a set of multi-component equilibrium models of
its disk. The dynamics of the disk is studied numerically using collisionless-gaseous numerical
simulations. After approximately one Gyr, a prominent central bar is formed with a semi-axis
of about three kiloparsecs. Outside the central regions, a multi-armed spiral
pattern develops, which can be characterized by the superposition of m=2, 3 and 4-armed
spiral patterns. The spiral structure and the bar exist for at least 3 Gyr in our simulations. The
presence of the bar in the disk of the MilkyWay galaxy imposes rather strict limitations on the
density distributions in the subsystems of the MilkyWay galaxy.We find that bar does not form
if the radial scale length of the surface density distribution of the disk is more than 2.6 kpc.
Analogously, the formation of bar is suppressed in the Milky Way disk in models with a massive
and compact stellar bulge. If future GAIA observations confirm the existence
of the three-four armed spiral pattern in the disk of the MilkyWay, this will prove the long-term
existence of spiral patterns in galactic disks.
 
\end{abstract}

\section{Introduction}           
\label{sect:intro}

Spiral structure in the Milky Way galaxy is subject of intensive research during many years (see,e.g., 
Dobbs $\&$ Baba 2014). Yet the morphology of the Milky Way spiral structure (number of arms, pitch angle, position(s) 
of corotation resonance(s) remain obscurred because we view the Galactic disk and its spiral structure edge on. 
Many attempts have been made to model the spiral structure in the disk-like galaxies.
Most of simulations employ either purely collisionless models (e.g., Fujii et al. 2018) or the hydrodynamical codes
(Renauld et al. 2013). Recently Grand et al.(2012) modeled the dynamics of Milky Way-sized galaxy 
using combined N-body-SPH hydrodynamical codes. They find the particular spirals are the recurrent transient features, continuously
reappearing in the numerical simulations. 

Transient feature of the particular spiral arms in the numerical simulations may be also
a consequence of the multi-armed nature of spiral structure, when disks are unstable towards a few global spiral modes with
different numbers of arms, pitch angles, angular velocities and amplitudes, so that a complex interplay between a few multi-armed
spirals leads to appearance and disappearance of the particular siral armlets.

We re-address the problem of the formation of spiral structure in the Milky Way-like galaxy using combined N-body -hydrodynamical code.
We employ the direct particle-particle integration scheme contrary to the approximate tree-code or the particle-mesh codes used in previous simulations.
Using equilibrium properties of the Milky Way disk such as its rotation curve,
velocity dispersion profiles and density distributions of the stellar and the gaseous components, we simulate the dynamics of a
multi-component stellar-gaseous disk and determine the number of spiral arms together with the perturbed velocity field 
and the lifetime of the spiral pattern. Comparison of the theoretically predicted properties of the spiral structure with observations
will enable to conclude whether the spiral pattern in the Milky Way disk is a long-lived phenomenon.

Korchagin et al. (2016) modeled the dynamics of a two-component stellar-gaseous Milky Way disk using 2D-simulations.
These authors found that a three-armed spiral pattern is generated in the Milky Way disk, and its spiral structure is sustained at
least 3 Gyr. We present here the results of three-dimensional simulations of the dynamics of a Milky Way stellar-gaseous disk. We model
the dynamics of perturbations using set of equilibrium models allowed within observational errors, and
demonstrate that in most of the models a multi-armed spiral pattern represented 
a superposition of spirals with different azimuthal wavenumbers is generated in the disk.

The Gaia mission will provide information about the
positions and kinematics for more than a billion stars in the disk of the Milky Way. These data will enable us to infer more reliable
conclusions about the morphology of our spiral structure in our Galaxy. The observational determination of the morphology of the spiral
pattern will also allow to shed light on another long-standing problem of galactic dynamics, i.e. it can help clarify whether 
the spiral structure in galaxies is a long-lived or a short-lived phenomenon. 

\section{Observational Data}

\label{sect:Obs}
Using COBE/DIRBE observational data (Drimmel \& Spergel 2001) found that the distribution of stars in the Milky Way disk can
be approximated by exponential function with radial scale length of
0.28 $R_\odot$, where $R_\odot$ is the distance of the Sun from the center of the Galaxy. 
Adopting the distance from the Sun to the center of Galaxy to be 8 kpc, we get a galactic disk radial scale length of 2.24 kpc.
Recent estimate of the radial scale length for Milky Way old stellar populations within $4 < r < 15$ kpc gives the value of 2.2 $\pm$ 0.2 kpc
(Bovy et al. 2016). From study of the density distribution of red clump stars outside the solar circle (Liu et al. 2017) find that
the radial scale length of the old stellar disk has a value of $r_d = 2.37 \pm 0.02 $ kpc. We assume that the density distribution of the
Milky Way old stellar disk along the radius is exponential with the scale length of 2.25 kpc.

Observations of edge-on galaxies show that the vertical density distribution of galactic thin disks can be approximated by 
the function
\begin{equation}\label{eq:VerticProfileDensity}
\varrho_{*}(r,z) = \varrho_{*0}(r)\, \textrm{sech}^{-2}(z/h_*) \,,
\end{equation}
where $h_*$ is the vertical scale height of the disk, and $\varrho_{*0}$ is the central volume density.
The stellar density distribution of the Milky Way galaxy is no exclusion from this rule. The three-dimensional distribution of 
old stellar populations of Milky Way can be represented by two subsystems: the thin disk
with a radial scale length about 2.5~kpc and a vertical scale height of 300 pc and a thick disk with a scale length and height 
of 900~pc and 3.6\,kpc respectively.

There is agreement between different determinations of the rotation curve of the galactic disk for $r\simeq 6\div 14$~kpc. 
The Galactic rotation curve has a local maximum in the inner disk regions as observed in atomic and molecular hydrogen lines. 
Outside the stellar disk the rotation curve of the Milky Way slowly decreases with radius reaching about 160 km\,s$^{-1}$ at distances of
$\simeq 100$~kpc from the center of the Galaxy.

Oservational data for the velocity dispersion of the Milky Way disk can be approximated by the exponential function.
A summary of data on observational determinations of the radial dependence of the velocity dispersion of stars $c_r$ in the
Milky Way gives the exponential scale length about 7.4~kpc, and the central velocity dispersion about 90 km/sec. 

Important parameter is the ratio of the velocity dispersions $c_z/c_r$. Recent estimate by Peng et al.(2018)
gives the ratio $c_z/c_r$ of 0.46.  We choose in our simulations the value of 0.5 which 
we use to build our equilibrium models. 

A number of observational studies demonstrate that the Milky Way has a boxy/peanut-shaped bulge. 
Estimates of the mass of the Milky Way bulge vary in a rather wide range.  
We vary the total mass of the bulge in our simulations within $(0.7 - 1.6)\times 10^{10}\, M_{\odot}$.

Gas plays a significant or even crucial role in the formation and sustaining the spiral structure in galaxies. 
The total amount of gas within 30~kpc of the Milky Way disk is about $8 \times 10^{9} M_{\odot}$ (Nakanishi $\&$ Sofue 2016).
In our simulations we vary the total amount of the disk gaseous component
from $3.5 \times 10^{9} M_{\odot}$ to $6.5 \times 10^{9} M_{\odot}$ within the $r=R_{opt}=4r_d=9$\,kpc as listed in Table~1.

\section{Numerical code and stability criteria}

We treat the dynamics of 3D stellar-gaseous galactic disk self-consistently . 
The gaseous subsystem is modeled by $N_g$ Smoothed Particle Hydrodynamics Particles (SPH), and the dynamics of the collisionless disk (stars) 
is modeled by using $N_*$ particles and the direct particle-particle integration scheme. 
We assume that the number of stellar and gaseous particles are equal $N_g = N_* = N/2$, where $N$ is the total number of particles 
used in the numerical simulations. For the numerical integration of the equations of motion we use the predictor-corrector scheme of 
second order accuracy. Details of the realization of the predictor-corrector method are described in papers by Khrapov $\&$ Khoperskov (2017).

Gravitating stellar-gaseous disks can be unstable to spiral perturbations. 
Toomre (1964) derived stability criterion that determines the growth of small-scale spiral perturbations:

\begin{equation}\label{eq:defQTstar}
Q_{T*} = \frac{c_r}{c_{T*}} \,, \quad c_{T*} = \frac{3.36 G \sigma_{*}  }{\kappa} \,,
\end{equation}

The galactic disks can be unstable towards the large-scale spiral perturbations (global modes) even when the 
local $Q$-stability parameter is more than unity, which is, in particular, the case  in our numerical simulations.
Figure 1 shows the radial dependence of Toomre Q-parameter for two models on our simulations. 
Disks of real galaxies are multi-component, and formulation of a local stability criterion for such systems is not an easy task. 
A few local stability criteria were suggested. We use the two-component criterion, suggested by Romeo $\&$ Falstad (2013) shown
in Figure 1 together with the local stability criteria for gaseous and for the collisionless components.

\begin{figure}[t]
	{\begin{center}
			\includegraphics[width=0.3\textwidth,angle=0]{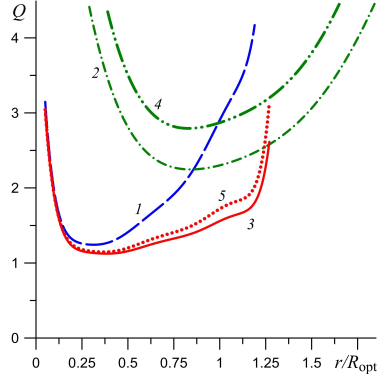}
			\caption{Radial dependence of Q-parameter depending on mass of gaseous 
				component. Curve 1 — Q-parameter of the stellar disk, 
				curve 2— Q parameter for the gaseous component (mass 4.1 $\times 10^9 M_{\odot}$),
				4 — Q parameter of gaseous disk with mass 3.9 $\times 10^9 M_{\odot}$. 
				Curves 3 and 5 - two-fluid Q parameter for these models
				correspondingly.}
			\label{f1}
	\end{center}}
\end{figure}

\section{Results}

Figure 2 shows the typical temporal evolution of the stellar gaseous disk with equilibrium parameters that agree with observational
data. The formation of bar in the central regions of the disk occurs rather quickly after approximately 1.5
disk revolutions. Outside the stellar bar, a complex spiral pattern grows which can be represented as a superposition of 2-3-4 spiral patterns 
of different amplitudes. The gaseous disk demonstrates even more complicated structure due to nonlinear interaction of the unstable
modes which leads to the branching of spirals, and appearance of the rows.

The number of spiral arms in the Milky Way galaxy, and their pitch angles has been the subject of discussions for a long time. 
Estimates of the pitch angle of the Milky Way spiral pattern range from $5^{\circ}$
to $25^{\circ}$.
The existing disagreement between different determinations of the properties of spiral arms in the Milky
Way based on observational data (masers, ionized. neutraland molecular hydrogen, 2MASS sources) is a manifestation of a
complex and non-stationary spiral structure in the Milky Way disk. The spiral structure of the Milky Way galaxy 
is, probably,  a superposition of nonlinear spiral patterns with different azimuthal wavenumbers, angular speeds of the patterns, and amplitudes.

\begin{figure}[t]
	{\begin{center}
			\includegraphics[width=0.71 \textwidth,angle=0]{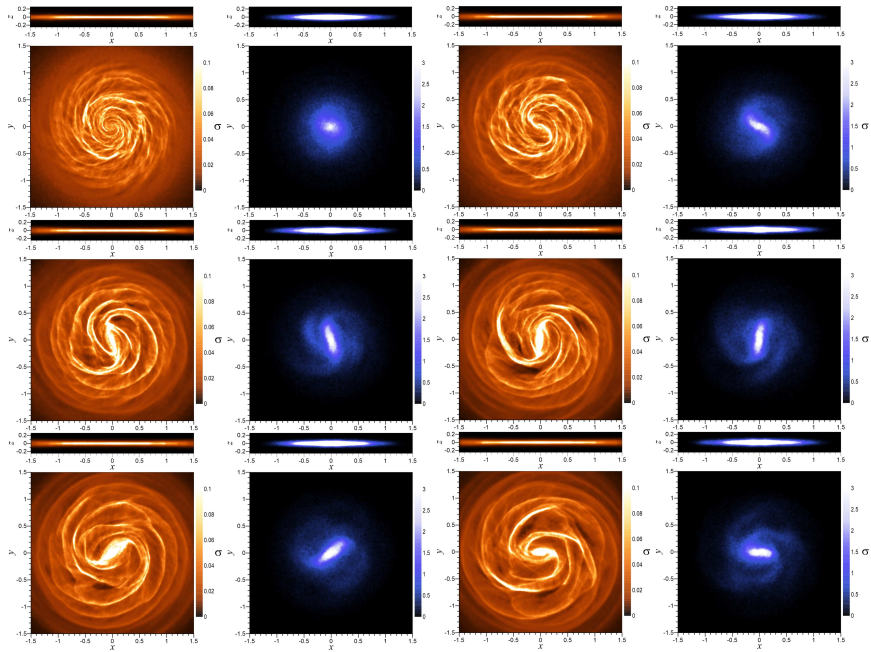}
			\caption{Snapshots of the density distributions in stellar (blue) and in gaseous (orange) components at different moments of time.}
			\label{f2}
		\end{center}
	}
\end{figure}

Due to presence of the bar, the kinematical properties of the stellar and of the gaseous disks inside $r\lesssim 3$ kpc demonstrate 
some specific features. Figure 3 shows the rotation curves of the disk along lines that have different angular 
orientations with respect to the semi-major axis of the bar. As one can see, the rotation curve of the disk along the semi-minor axis of 
the bar has a local maximum inside 1 kpc from the center of galaxy, similar to the observed kinematical properties of the Milky Way 
disk within the central kiloparsec, and along semi-major axis such kinematical feature is absent.

\begin{figure}[t]
	{\begin{center}
			\includegraphics[width=0.4\textwidth,angle=0]{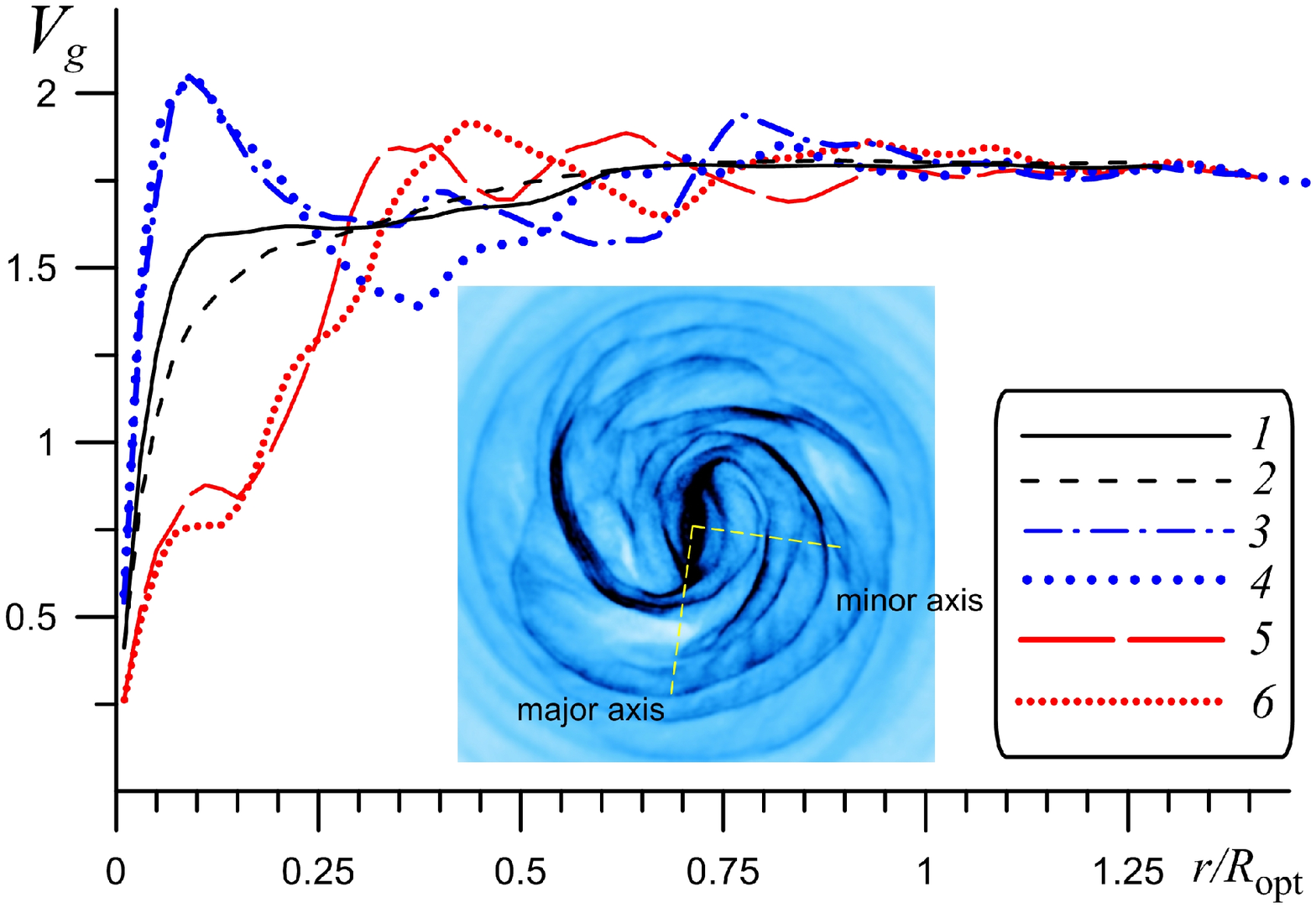}
			\includegraphics[width=0.4\textwidth,angle=0]{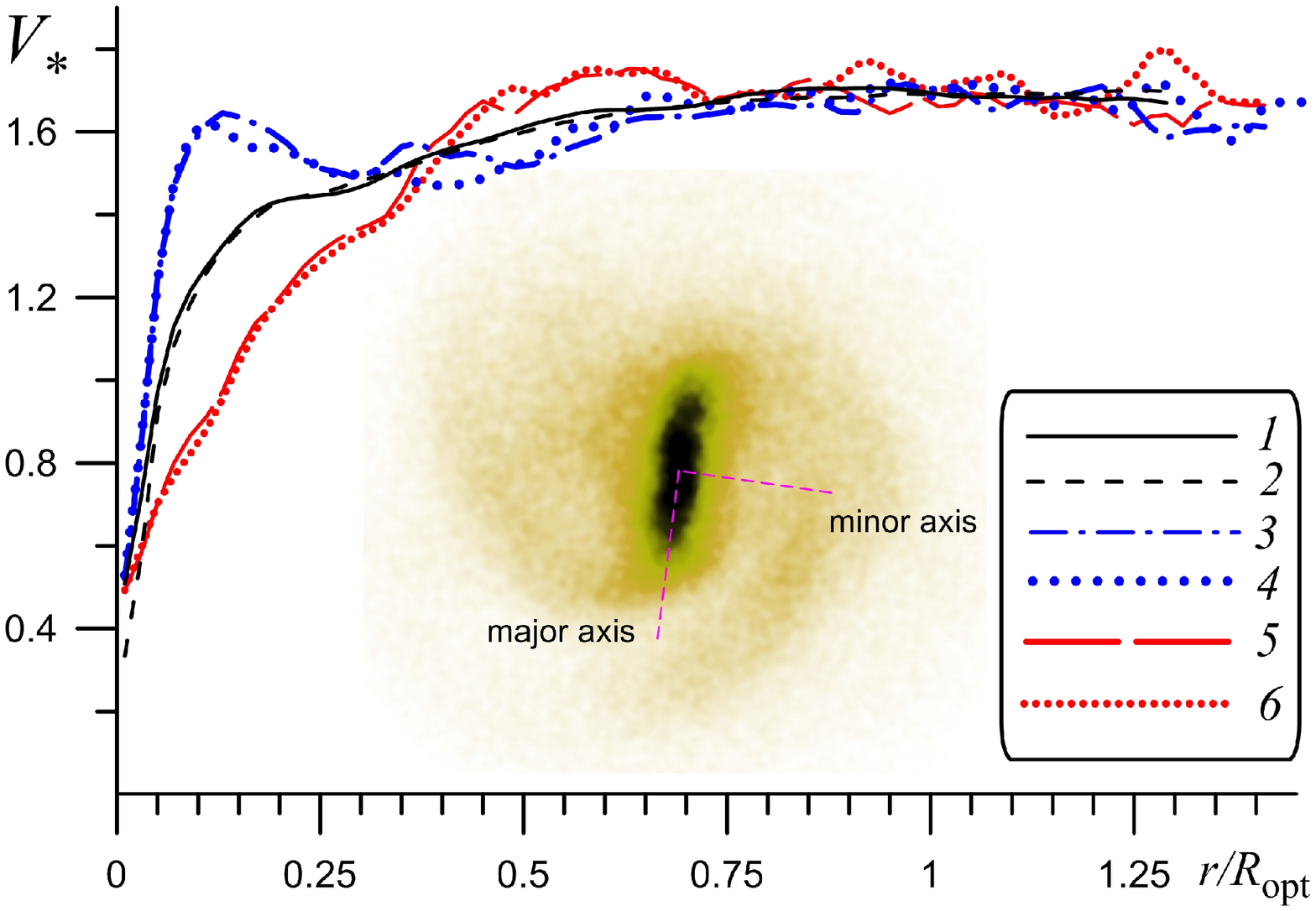}
			\caption{Typical profiles of the azimuthal velocities of the stellar disk
				(right frame) and the gaseous disk (right frame) along the small (curves 3
				and 4) and large (curves 5 and 6) axis of the bar. 
				For the comparison, the azimuthally averaged velocity profiles
				are also shown. Curve 1 shows the azimuthal velocity at t = 0. The approximate size of the bar
				is also shown.}
			
		\end{center}
	}
\end{figure}

Formation of bar is sensitive to the relative mass of the dark matter halo $\mu$, and to the mass 
of the bulge. In models with moderate masses of halo $\mu\simeq 1.6$, but with a massive 
and compact bulge bar is not developed. Thus, presence of bar in the Milky Way disk puts limitations on its equilibrium 
properties.

\section{Summary}
\label{sect:conclusion}

We demonstrate that the azimuthally averaged disk with the parameters of the Milky Way galaxy, namely, with the observed rotation curve, 
velocity dispersion profile, and with masses of bulge, of stellar and gaseous disk components that agree with the available observational data, 
is unstable towards the growth of a number of spiral modes. At the nonlinear stage of instability, the perturbations saturate at the level
from tens in central regions, to a few percent at disk periphery. At the saturation stage, a prominent bar is formed in the 
central regions of the disk with large semi-axis of about 3 kpc. Outside bar region, a complex spiral structure, represented by a  
superposition of two-, three-, and four-armed spiral patterns, rotating with different angular velocities, is formed. 
Traditionally, the observational data of spiral structure in the Milky Way galaxy are interpreted in terms of the number of arms, 
pitch angle and a position of the co-rotation resonance for one spiral mode.
Our simulations show that the spiral pattern in the Milky Way disk cannot be described in terms of one 
mode with fixed number of arms and fixed pitch-angle.

Another main conclusion, made from our simulations, is that spiral structure in the Milky Way disk is long-lived. 
The spiral structure evolves with time, but spiral pattern lasts more than 3 Gyr. The spiral armlets appear and disappear due-to
interference of spiral patterns with different numbers of arms and with different angular velocities.

We demonstrate also that peak on the rotation curve of the Milky Way disk, observed in its central regions, 
is a result of the formation of bar, and cannot be explained by presence of a massive and 
centrally concentrated bulge. Models with such bulge prevent the formation of the bar in the disk, and contradict observational data.


\section*{References}

\begin{thebibliography}{9}
	
	\bibitem {Dobbs} Dobbs, C., Baba, J. 2014, PASA, 31, 1
	\bibitem {Fujii} Fujii, M., Bedorf, J., Baba, J., Zwart, P. 2018, MNRAS, 477, 1451
	\bibitem {Grand} Grand, R., Kawata, D., Cropper, M. 2012, MNRAS, 426, 167
	\bibitem {Khrapov} Khrapov, S., Khoperskov, A. 2017, Communications in Computer and InInformation Science, 793, 266
	\bibitem {Korchagin} Korchagin, V.I., Khoperskov, S.A., Khoperskov, A.V. 2016, Baltic Astronomy, 25, 356
	\bibitem {Nakanishi} Nakanishi, H., Sofue, Y. 2016, PASJ, 68, 5
	\bibitem {Peng} Peng, X., Wu, Z., Qi, Z. et al. 2018, PASP, 130, 074102
	\bibitem {Renaud} Renaud, F., Bournaud, F., Emsellem, E. et. al. 2013, MNRAS, 436, 1836
	\bibitem {Romeo} Romeo, A. B., Falstad, N. 2013, MNRAS, 433, 1389
	\bibitem {Toomre} Toomre, A. 1964, ApJ, 139, 1218
	
\end{thebibliography}
\end{document}